# The Rôle of Diffraction in the Quantization of Dispersing Billiards

Harel Primack[1], Holger Schanz[2], Uzy Smilansky[1] and Iddo Ussishkin[1]

[1]*Department of Physics of Complex Systems, The Weizmann Institute of Science, Rehovot 76100, Israel*

[2]*Institute of Physics, Humboldt-University, Berlin, Germany*

(August 9, 1995)

## Abstract

We study diffraction corrections to the semiclassical spectral density of dispersing (Sinai) billiards. They modify the contributions of periodic orbits (PO's), with at least one segment which is almost tangent to the concave part of the boundary. Given a wavenumber $k$, all the PO's with length up to the Heisenberg length $O(k)$ are required for quantization. We show that most of the contributions of PO's which are longer than a limit $O(k^{2/3})$ must be corrected for diffraction effects. For orbits which just miss tangency, the corrections are of the same magnitude as the semiclassical contributions themselves. Orbits which bounce at extreme forward angles give very small terms in the standard semiclassical theory. The diffraction corrections increase their amplitude substantially.

05.45.+b, 03.65.Sq

Dispersing (Sinai) billiards were the first class of billiards for which classical chaoticity was rigorously proven [1]. These billiard are defined by boundaries which consist of concave circular arcs and possibly neutral (straight) segments. Chaos is generated by the divergence of neighboring orbits when they bounce off the concave arcs. Whenever a classical trajectory is tangent to an arc, the classical billiard map is discontinuous. This singularity in the classical dynamics introduces diffraction effects when the billiards are semiclassically quantized. Due to the finite wavelength, the domain which is affected consists of a finite neighborhood of the classical tangent trajectories, and is called the *penumbra* (almost shadow). In the present paper we shall analyze the effect of diffraction in the penumbra on the semiclassical quantization of dispersing billiards. We shall show that it introduces modifications to the standard semiclassical trace formula [2], which are of the same order as the leading terms. Moreover, it will be shown that diffraction effects are significant because classical PO's which are relevant for semiclassical quantization traverse the penumbra with very high probability.

Diffraction corrections in the penumbra are to be distinguished from contributions of creeping trajectories, which are described by the geometrical theory of diffraction [3]. Creeping corrections were previously included in the trace formula [4], and were shown to be exponentially small. Thus, the present theory applies to orbits which fall outside the domain of validity of both the standard semiclassical treatment, and its extension which includes creeping orbits.

For the sake of clarity of presentation, our analysis will concentrate on the simplest dispersing billiard, whose boundary $\Gamma$ consists of a square $S$ and a concentric inscribed circle $C$. We shall refer to it as the Sinai billiard. We introduce the diffraction effects by using a variant of the boundary integral method: The eigenvalues of the billiard are those real values of $k$ for which the boundary integral equation

$$u(\mathbf{r}_s) = 2 \int_S \mathrm{d}s' \frac{\partial G_c}{\partial \hat{n}_s}(\mathbf{r}_s, \mathbf{r}_{s'}) u(\mathbf{r}_{s'}) \tag{1}$$

has a solution. In contrast to the standard boundary integral method, we use the *circle's* Green function $G_c(\mathbf{r}, \mathbf{r}')$, and the integration is carried out along the boundary *excluding* the circle, i.e. along $S$. The circle's Green function satisfies $(\Delta + k^2)G_c(\mathbf{r}, \mathbf{r}') = -\delta(\mathbf{r} - \mathbf{r}')$ in the exterior of $C$, with Dirichlet boundary conditions on $C$ and outgoing boundary conditions at infinity.

The contribution of PO's which bounce $N$ times off $S$ to the density of states $d(k)$, is derived from (1), by considering the $N$'th term in the multiple reflection expansion

$$\text{Im} \frac{2^N}{\pi N} \frac{d}{dk} \int_S ds_1 \ldots ds_N \frac{\partial G_c}{\partial \hat{n}_1}(\mathbf{r}_1, \mathbf{r}_2) \cdots \frac{\partial G_c}{\partial \hat{n}_{N-1}}(\mathbf{r}_{N-1}, \mathbf{r}_N) \frac{\partial G_c}{\partial \hat{n}_N}(\mathbf{r}_N, \mathbf{r}_1). \tag{2}$$

The PO contributions are identified when these integrals are evaluated by the stationary phase approximation, with each saddle point corresponding to one PO. For this purpose $G_c(\mathbf{r}, \mathbf{r}')$ should be approximated by a sum of terms of the form $Ae^{ikL(\mathbf{r}, \mathbf{r}')}$, where $A$ varies smoothly with $\mathbf{r}$ and $\mathbf{r}'$.

The circle's Green function is expressed exactly, in polar coordinates, by an angular momentum expansion and by Poisson resummation, as $G_c(\mathbf{r}, \mathbf{r}') = \sum_{m=-\infty}^{\infty} G_c^{(m)}(\mathbf{r}, \mathbf{r}')$, where

$$G_c^{(m)}(\mathbf{r}, \mathbf{r}') = \frac{i}{8} \int_{-\infty}^{\infty} dl \left( H_l^-(kr_<) + S_l(kR) H_l^+(kr_<) \right) H_l^+(kr_>) e^{il\Delta\theta + 2\pi i m l}. \tag{3}$$

In this equation, $r_>$ ($r_<$) is the larger (smaller) of $r$ and $r'$, $S_l(kR) = -H_l^-(kR)/H_l^+(kR)$ is the diagonal element of the circle scattering matrix, and the angle difference $\Delta\theta = |\theta - \theta'|$ is always taken so that $0 \leq \Delta\theta \leq \pi$. We assume that $r, r' \gtrsim R + R(kR)^{-1/3}$, i.e. the two points are not in the near vicinity of the circle. The term $G_c^{(0)}(\mathbf{r}, \mathbf{r}')$ gives the dominant contribution. We shall briefly outline the derivation of the standard semiclassical and creeping results, and then, the leading corrections in the penumbra will be derived.

The domain of $\mathbf{r}$ and $\mathbf{r}'$ in which the standard semiclassical result holds is called the *illuminated* region (see Fig. 1(a)). In this region, the integral for $G_c^{(0)}(\mathbf{r}, \mathbf{r}')$ is evaluated using the Debye approximation for the Hankel functions [5], and the integration is performed using the stationary phase approximation. There are two saddle points, which relate to the two classical trajectories from $\mathbf{r}$ to $\mathbf{r}'$: One is direct, and the other reflects once from the circle. Using the two contributions in (2), one recovers the standard semiclassical trace formula. The Debye approximation for $S_l(kR)$ fails if $(kR - l_r) \lesssim (kR)^{1/3}$, where $l_r$ is the angular momentum of the reflected trajectory. This sets the limit of validity of the semiclassical approximation, and defines the borderline between the illuminated region and the penumbra.

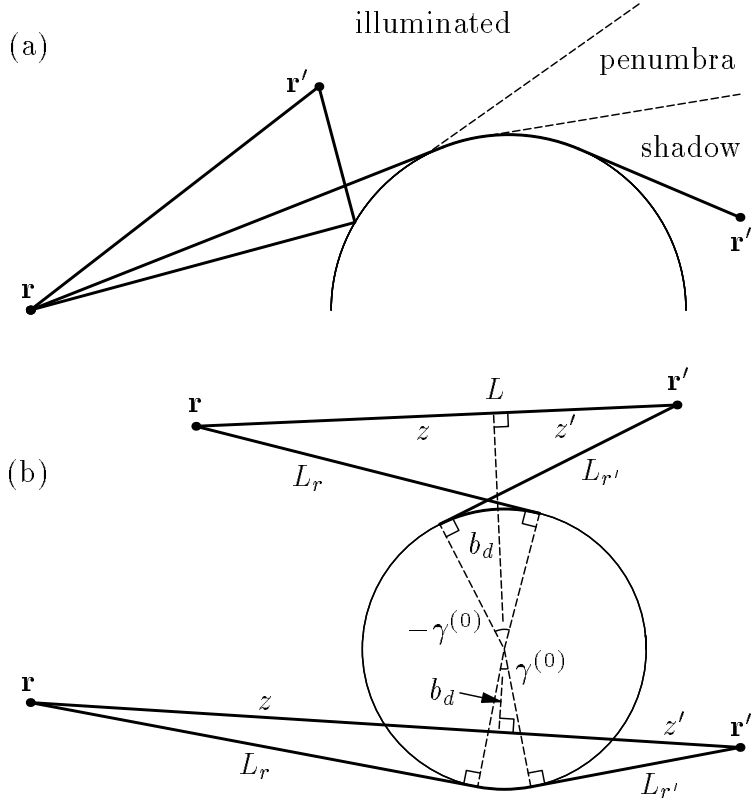

FIG. 1. (a) The three regions of $\mathbf{r}'$ for a fixed $\mathbf{r}$. The direct and reflected trajectories are shown for $\mathbf{r}'$ in the illuminated region, and the shortest creeping trajectory is shown in the shadow region. (b) The geometrical setup in the penumbra. In the upper part $\mathbf{r}$ and $\mathbf{r}'$ are in the classically illuminated region ($b_d > R$), and in the lower part they are in the classically shadowed region.

The term $G_c^{(0)}(\mathbf{r},\mathbf{r}')$ gives the exponentially small contribution of a creeping trajectory in the *shadow* region (see Fig. 1(a)). It is obtained by closing the contour of integration in (3) in the complex $l$ plane, and summing over the poles of $S_l(kR)$ [3]. Using creeping contributions in (2) one recovers the results of [4]. This procedure fails if $\gamma^{(0)} \lesssim (kR)^{-1/3}$, where $\gamma^{(0)}$ is the creeping angle of the trajectory. The above condition defines the borderline between the shadow region and the penumbra. A similar evaluation of the terms $G_c^{(m)}(\mathbf{r},\mathbf{r}')$ with $m \neq 0$ results in creeping contributions for any $\mathbf{r}$ and $\mathbf{r}'$.

In the penumbra, neither of the above approximations hold, and one should use specially designed methods for approximating $G_c(\mathbf{r},\mathbf{r}')$, such as the one introduced by Nussenzveig [6]. The expression for $G_c^{(0)}(\mathbf{r},\mathbf{r}')$ is split into two parts, a *direct* part and a *glancing* part, which are respectively defined by

$$G_d(\mathbf{r},\mathbf{r}') = \frac{i}{8}\int_{kR}^{\sigma_2\infty} dl\, H_l^+(kr)H_l^+(kr')e^{il\Delta\theta}, \tag{4a}$$

$$G_g(\mathbf{r},\mathbf{r}') = \frac{i}{8}\int_{\sigma_1\infty}^{kR} dl\, S_l(kR)H_l^+(kr)H_l^+(kr')e^{il\Delta\theta}$$
$$+ \frac{i}{8}\int_{kR}^{\sigma_2\infty} dl\, (S_l(kR) - 1)H_l^+(kr)H_l^+(kr')e^{il\Delta\theta}. \tag{4b}$$

The limits $\sigma_1\infty$ and $\sigma_2\infty$ are directed infinities in the complex plane, defined in [6], such that the full contour from $\sigma_1\infty$ to $\sigma_2\infty$ goes through $kR$ and around all the poles of $S_l(kR)$.

The evaluation of the direct part is similar to that of the direct trajectory contribution in the illuminated region. The only difference is that here we take into account the boundary of the integral at $kR$. As a result, the semiclassical contribution of the direct trajectory is multiplied by the factor $(F(\infty) - F(\nu))/\sqrt{2i}$, where $F(x) = C(x) + iS(x)$ is the Fresnel integral function, and

$$\nu = \left(\frac{kL}{\pi z z'}\right)^{\frac{1}{2}} (R - b_d). \tag{5}$$

The impact parameter of the direct path is denoted by $b_d$, $z = \sqrt{r^2 - b_d^2}$, $z' = \sqrt{r'^2 - b_d^2}$ and $L = z + z'$ is the length of the direct trajectory (see Fig. 1(b)). The Fresnel factor equals $\frac{1}{2}$ for exact tangency, and goes to 1 (0) when approaching the borderline of the illuminated (shadow) region.

The main contribution to the glancing part (4b) comes from the vicinity of $l = kR$. In the integrand, $S_l(kR)$ is replaced by its transition region approximation for $l \approx kR$ [5]. The rest of the integrand is evaluated at $l = kR$, where the Debye approximation is used for the Hankel functions. This way, one obtains

$$G_g(\mathbf{r},\mathbf{r}') = -\frac{C}{4\pi} \frac{(kR)^{1/3}}{k\sqrt{L_r L_{r'}}} e^{ik(L_r + L_{r'} + R\gamma^{(0)})}, \tag{6}$$

where $C \approx 0.996193019928 e^{i\pi/3}$ was obtained by a numerical integration. The length of the line segment from $\mathbf{r}$ ($\mathbf{r}'$), to the point where it is tangent to the circle, is $L_r$ ($L_{r'}$), and $\gamma^{(0)}$ is the directed creeping angle (see Fig. 1(b)). When $b_d < R$ the length $L_r + L_{r'} + R\gamma^{(0)}$ is equal to the length of a creeping path from $\mathbf{r}$ to $\mathbf{r}'$. When $b_d > R$ the angle $\gamma^{(0)}$ becomes negative.

The contribution of orbits that traverse the penumbra to the density of states $d(k)$ may now be found, using (2) and the appropriate approximation for the Green function in each of the segments of the orbit. For isolated PO's that traverse the penumbra once, with a direct segment, the standard semiclassical contribution (which is $O(k^0)$) is multiplied by the Fresnel factor. Thus, the correction to the trace formula will be of the same order as the contribution itself. In the case of a single glancing traversal in the penumbra, the contribution is proportional to $k^{-1/6}$. Due to the new form of the phase and the prefactor of $G_g(\mathbf{r}, \mathbf{r}')$, there is no direct relation between the standard semiclassical result and the modified one. The former is very small since at glancing, the classical motion is extremely unstable. This is manifested in the standard semiclassical result by the fact that $G_r(\mathbf{r}, \mathbf{r}')$ is very small for an almost tangent reflection. Replacing it with $G_g(\mathbf{r}, \mathbf{r}')$, which has a larger amplitude, the contribution of the PO is typically enhanced. The extension to PO's with few segments that traverse the penumbra is straightforward. Since $G_c(\mathbf{r}, \mathbf{r}')$ gives a contribution in the classically shadowed region of the penumbra, there will also be significant contributions of PO's with classically forbidden segments.

We shall present now a few examples which illustrate the importance of the penumbra corrections. This is well demonstrated in terms of the length spectrum

$$D(x) = \int_0^\infty \mathrm{d}k\, w(k) e^{ikx} d(k), \tag{7}$$

where $w(k)$ is a weight function. Every PO contributes to $D(x)$ in a small vicinity of its length, and this allows us to isolate the contribution of a single PO (or of few orbits with similar lengths) and to compare it to the theory.

We consider the quarter Sinai billiard (see Fig. 2), with side 1 and radius 0.5. The exact

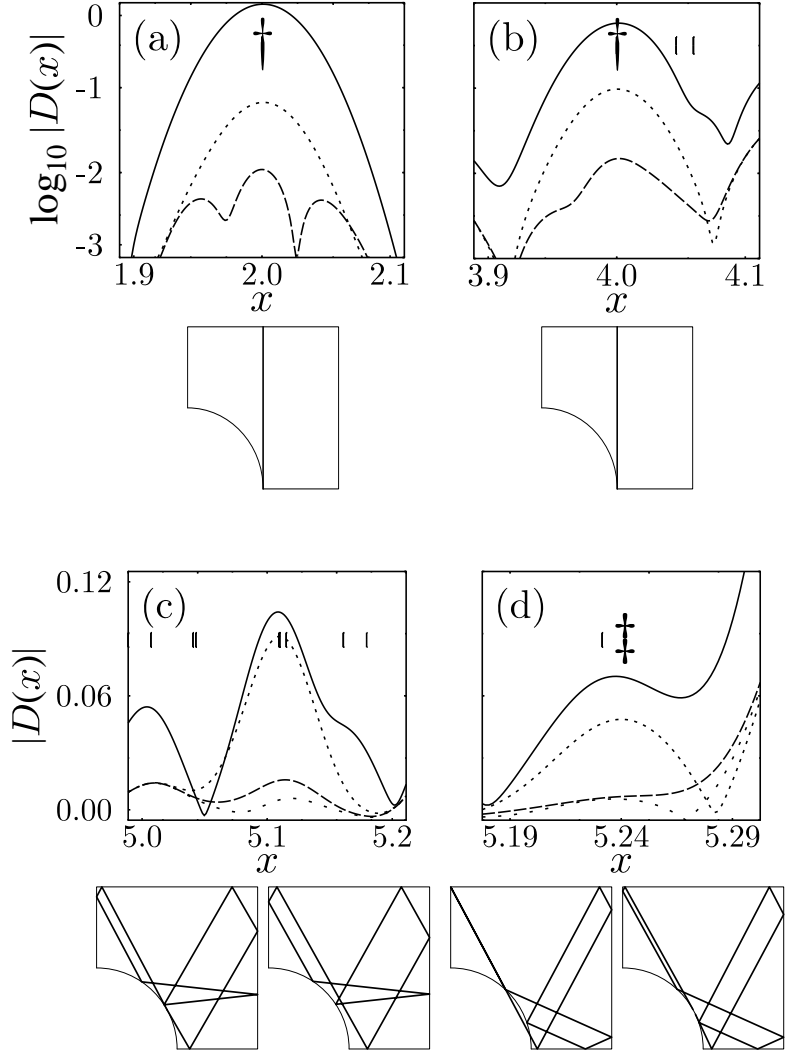

FIG. 2. Penumbra corrections of the length spectrum for four selected cases. Solid lines show the quantum (exact) length spectrum. All other lines show the *deviations from* various semiclassical approximations (SCA). The orbits considered are shown below each frame. Vertical bars indicate locations of unstable PO's, daggers indicate bouncing ball families. (a) Shortest tangent orbit ($x = 2$). Dotted line - SCA including bouncing ball and edge contributions according to [10]. Dashed line - glancing contribution also included. Note the logarithmic scale. (b) Double traversal of the orbit considered in (a). Dotted line includes 3 penumbra contributions (see text). (c) Pair of almost tangent PO's at $x \approx 5.10$. Dotted line - standard SCA, dashed line - direct term included, sparse dotes - glancing contribution also included. (d) Pair of classically forbidden PO's at $x \approx 5.24$, indicated by double dagger. Notation is as in (c).

(quantum mechanical) eigenvalues in the range $0 \leq k \leq 300$ (total of 5667) were obtained using the scattering method described in [7]. The classical PO's were calculated using the minimum and the unique coding principles [8,9]. To obtain the results for the trace formula, we used the four-fold desymmetrization of the circle's Green function. The weight function $w(k)$ in (7) was taken as a Gaussian centered around $k_0 = 150$, and whose width is $\sigma = 40$.

We first consider exactly tangent orbits. They always exist at the edge of the one-parameter families of neutral PO's ("bouncing balls") that bounce only between the straight segments of the billiard. The simplest of these families is of length 2 (see Fig. 2(a)). According to the analysis in [10], the semiclassical contribution consists of two terms: One that describes the contribution of the family (and is $O(k^{1/2})$), and the other that takes into account the effect of the edge of the square (and is $O(k^0)$). Comparison with the exact length spectrum shows a significant difference. Application of our analysis in the penumbra, shows that in addition to the family and edge contributions, there is a term $O(k^{-1/6})$, coming from an isolated exactly tangent orbit. This term is obtained from $G_g(\mathbf{r}, \mathbf{r}')$. The additional contribution successfully accounts for the error of the semiclassical approximation.

The semiclassical result for the double repetition of this family is similar to the single repetition, and again shows a significant deviation from the exact result (Fig. 2(b)). Application of our analysis in this case gives three diffraction terms of orders $O(k^0)$, $O(k^{-1/6})$ and $O(k^{-1/3})$, which reduce the error substantially.

The case of unstable isolated PO's traversing the penumbra, is demonstrated for the two PO's of length $\approx 5.10$ (see Fig. 2(c)). The two orbits are geometrically similar, except that one (the direct) passes very close to the circle, and the other (the glancing) reflects in an extremely forward direction. The semiclassical amplitude of the direct orbit is multiplied by a Fresnel factor, which is $0.71 \exp(-0.23i)$ for $k = k_0 = 150$. This accounts for most of the semiclassical error. Including the corrected contribution of the glancing orbit, reduces the remaining difference by half.

Next, we illustrate the corrections due to classically forbidden PO's which traverse the shaded part of the penumbra (see Fig.1(b)). The two orbits shown in Fig.2(d) serve this pur-pose. Their length is $\approx 5.24$, one of them cuts through the disc and the other creeps around it. (If we were to reduce the radius of the disc continuously, these orbits would coalesce at $R \approx 0.48$ and appear as a single tangent orbit). The diffraction contributions of these non classical orbits successfully account for the large difference between the exact quantum length spectrum and its semiclassical counterpart at the relevant length. Application of the geometrical theory of diffraction [4] (including many modes) results in large deviations from the quantum mechanical length spectrum.

The above examples illustrated in detail the effect of diffraction corrections on the contributions from particular PO's. In the remaining, we provide some arguments showing that most of the relevant PO's are affected. One should bear in mind that the borderlines of the penumbra are $k$ dependent. Any orbit (which is not exactly tangent), is excluded from the penumbra for a sufficiently high wavenumber. Hence, the standard semiclassical contribution of any particular PO is valid in the semiclassical limit $k \to \infty$. Moreover, the fraction of phase space occupied by the penumbra, is of order $(kL)^{-2/3}$, where $L$ is a typical length of the billiard, and this fraction vanishes as $k \to \infty$. Thus one may naïvely conclude that the global effect of diffraction is negligible, which is not true because of the following reasons: For an orbit with $n$ segments, define $\Delta = \min_j |l_j - kR|$, where $l_j$ is the angular momentum of the $j$th segment with respect to the circle's center. The orbit traverses the penumbra (at least once) if $\Delta \lesssim (kR)^{1/3}$. Each segment of the orbit has an a-priori probability $p \approx (kL)^{-2/3}$ to traverse the penumbra. Assuming statistical independence of the segments, and homogeneous coverage of phase space by long PO's, the probability that the orbit will avoid the penumbra is $(1-p)^n \approx \exp(-n(kL)^{-2/3})$. In order to achieve an energy resolution of the order of the mean level spacing, one has to include in the semiclassical theory all the PO's whose lengths are smaller than the Heisenberg length $L_H \approx kL^2$. An orbit of this length has $n_H \approx kL$ segments. Thus, for all orbits with $n_H^{2/3} \lesssim n \lesssim n_H$ the probability that the orbit does not traverse the penumbra is very small. The exponential proliferation of PO's guarantees that most of the relevant orbits are in this range. Thus, simple ergodic considerations give the surprising result, that the semiclassical approximation fails for the *majority* of relevant PO's in the semiclassical limit $k \to \infty$. The contributions of orbits whose length is shorter than $\approx n_H^{2/3}$ are described by the standard semiclassical theory with a high probability. As $k \to \infty$, the number of these orbits grows, but they comprise a smaller fraction of the set of relevant PO's.

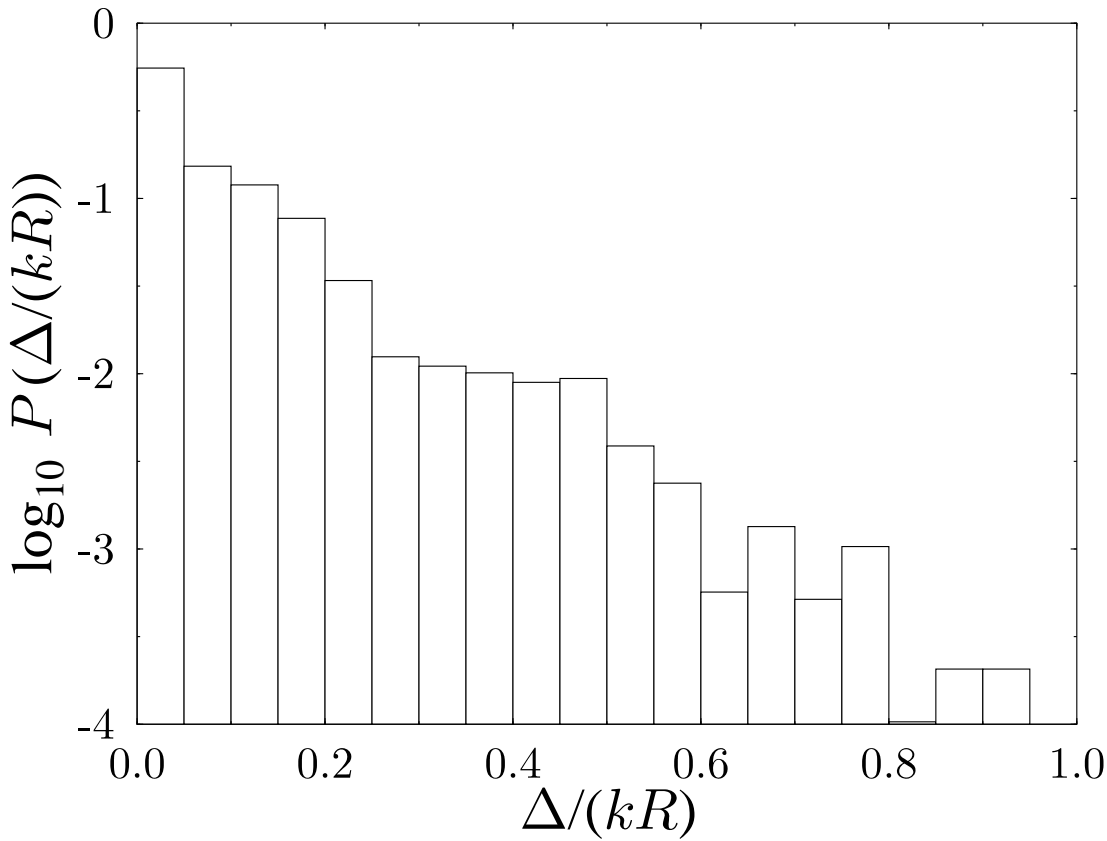

FIG. 3. The coarse-grained distribution of $\Delta/(kR)$, defined in the text. Only angular momenta $l_j$ of reflections from the circle are considered. The data is based on 19375 PO's which total to 107029 reflections from the circle. Note the logarithmic scale.

In Fig. 3 we present the distribution of $\Delta$ for all the PO's in the quarter Sinai billiard for lengths between 7 and 10. The distribution is sharply peaked at the minimal value, indicating that indeed most PO's include an almost tangent chord.

The results introduced in this Letter clearly show that diffraction corrections may play a very important rôle in the semiclassical quantization of dispersing billiards. There are two complementary factors which make the diffraction corrections in the penumbra significant:

First, the correction of amplitudes of individual orbits is of the same order as the semiclassical contribution itself, and hence must be considered prior to higher order corrections (e.g. [11]). Second, the corrections should be applied to a large fraction of the relevant PO's. The effect of diffraction corrections on an individual energy level, is still to be investigated.

This work was supported by the US-Israel Binational Science Foundation (BSF) and by the Minerva Center for Nonlinear Physics of Complex Systems. HS thanks the Weizmann Institute for its kind hospitality, and acknowledges support from the Deutsche Forschungsgemeinschaft.